%% file: ISIT11_capacity.tex
\documentclass[conference]{IEEEtran} 
\usepackage{amssymb}
\usepackage[cmex10]{amsmath}

\usepackage{graphicx} 

\usepackage{array, cite} 
\usepackage{subfigure,epsfig} 

\title{ Capacity of Discrete Molecular Diffusion Channels}
\author{ Arash Einolghozati, Mohsen Sardari, Ahmad Beirami, Faramarz Fekri\\
School of Electrical and
Computer Engineering, Georgia Institute of Technology, Atlanta, GA 30332\\
\texttt{Email:}\{einolghozati, mohsen.sardari, beirami, fekri\}@ece.gatech.edu}

\begin{document}

\maketitle

\begin{abstract}
In diffusion-based molecular communications, messages can be conveyed via the variation in the concentration of molecules in the medium. In this paper, we intend to analyze the achievable capacity in transmission of information from one node to another in a diffusion channel. We observe that because of the molecular diffusion in the medium, the channel possesses memory. We then model the memory of the channel by a two-step Markov chain and obtain the equations describing the capacity of the diffusion channel. By performing a numerical analysis, we obtain the maximum achievable rate for different levels of the transmitter power, i.e., the molecule production rate.
\end{abstract}
\section{Introduction}
\input{intro}

\section{Background}
\label{sec:back}

The first step is to characterize the temporal and spatial variations of molecules in the channel which follows the general diffusion equations. According to Fick's second law of diffusion the concentration of molecules $c(x,t)$ at position $x$ at time $t$ is computed using the molecule production rate $r(x,t)$, as follows:
\begin{equation}
\label{eq:eq4}
 \frac {\partial c(x,t)}{\partial t} = D \nabla^2 c(x,t)+r(x,t),
\end{equation}
Here, $x$ is the distance of any point in the environment from the source and $D$ is the diffusion coefficient of the medium. The impulse response of~(\ref{eq:eq4}), corresponding to $r(x,t)=\delta(x)\delta(t)$, is the Green's function $g_d(x,t)$ whose expression is as follows:
\begin{equation}
\label{eq:eq5}
 g_d(x,t)=\frac {1}{4\pi Dt} \exp{\left(-\frac{|x|^2}{4Dt}\right)}.\
\end{equation}
This impulse response is given for the 2-D medium but can be extended to a general 3-dimensional case, using the observation that $n$-dimensional diffusion is equivalent to $n$ separate (simultaneous) 1-D diffusions. Since the diffusion equation is a linear equation, the solution to ~(\ref{eq:eq4}) for an arbitrary input $r(x,t)$, denoted by $c^*(x,t)$, can be obtained using  
\begin{equation}
\label{eq:eq6}
 c^*(x,t)=g_d(x,t)\otimes r(x,t),
\end{equation}
where $\otimes $ denotes a multi-dimensional convolution operation on $x$ and $t$.

In our setup, we assume that there is only one transmitter emitting molecules. Therefore, in~(\ref{eq:eq6}), we have  $r(x,t)=F(t)\delta(x)$, where $F(t)$ is the input signal. Hence, we will have
\begin{equation}
\label{eq:eq7}
c^*(x,t) =\int^{\infty}_0 F(\tau)\frac {1}{4\pi D(t-\tau)} \exp{\left( -\frac{x^2}{4D(t-\tau)}\right)} \, \mathrm{d}{\tau}.
\end{equation}
This response is valid for open free media in which the only boundary conditions are at the transmitter. Note that in this model, we do not consider the delay due to the travel time of molecules between the transmitter and the receiver. That is we assume molecules reach the receivers instantly. This assumption, however, does not affect our analysis of the capacity because it only shifts the time that molecules are arrived at the receiver.

\section{Model}
\label{sec:model}

We consider the case of a single transmitter and a single receiver in a 2-D environment. This model can be extended to a network of transmitters and receivers in which each transmitter intends to communicate with a specific receiver when interference among these communications is negligible. We consider a discrete-time model for the communication in which an arbitrary string of binary information is given to the transmitter. This string is to be conveyed to the receiver at distance $r$ from the transmitter. 


The transmitter encodes the information bits into the proper concentration of molecules and releases them for some specific durations, discussed later. These molecules diffuse in the medium. The receiver compares the concentration of received molecules with a specific threshold and decides whether the concentration in the environment is low or high.
This threshold depends on the characteristics of the environment. It can neither be arbitrarily small because of the noise and interference in the environment nor be arbitrarily large because of the limited capability of the transmitter for molecule production. In the latter case, either the concentration may not reach the threshold or it would take a long time.

The main difference here with a typical paradigm of communication is the memory which essentially exists in the diffusion channel and influences the communication dramatically. The concentration of molecules cannot change momentarily and unlike the typical binary symmetric channel, previously sent symbols affect the new ones. For example, suppose the concentration of molecules in the channel is in the high state and the transmitter intends to send ``0'' through the channel. However, molecules from the previous state still exist in the environment and cannot be removed suddenly. Hence, the transmitter has to wait a specific amount of time to send the new bit. Note that we do not assume existence of negative rates because it is unrealistic to assume as such for the molecular communication. We also do not consider any noise to simplify the study.

\begin{figure}
\centering
\vspace{-0.05in}
\includegraphics[width = .4\columnwidth, angle = -90]{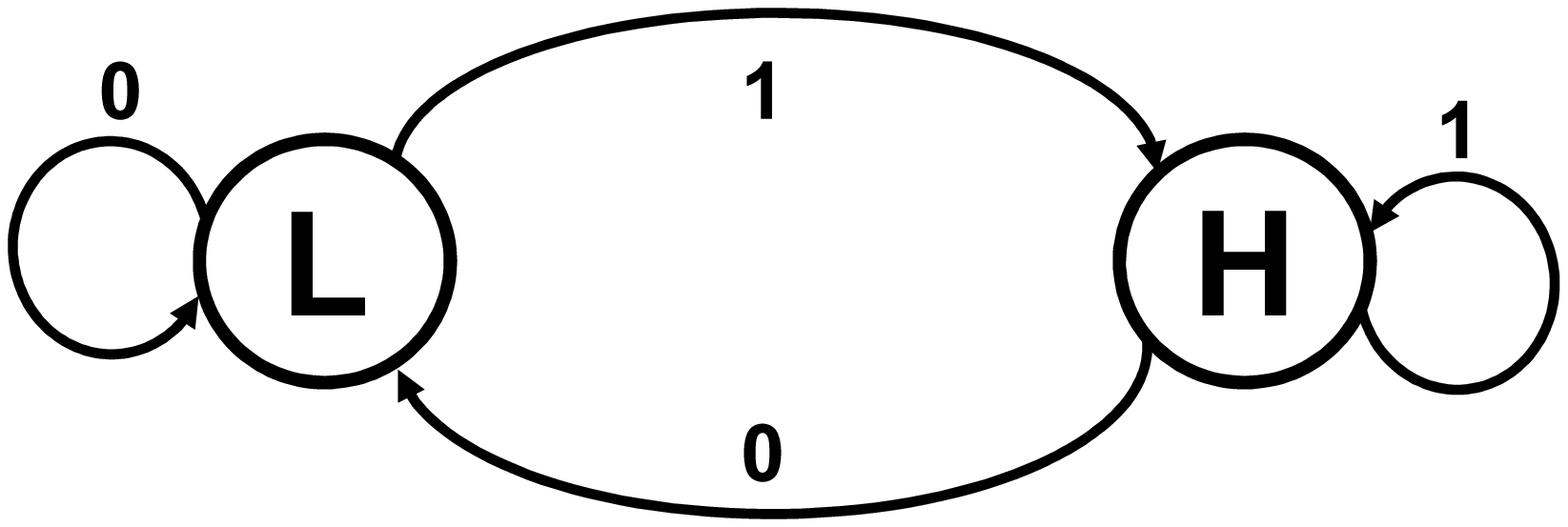}
\caption{Two states of the channel and possible inputs in each state.}
\vspace{-0.15in}
\label{fig:states}
\end{figure}

Based on the previous discussion, we observe that each bit depends only on the last bit that was transmitted and it is independent of the other previously transmitted bits. Therefore, a Markov chain can be used to model the communication in the diffusion channel. In such a model, the state of the channel is set by the last transmitted bit: the channel state is ``H'' if the last transmitted bit was ``1'', and it is ``L'' otherwise. Hence, there are two states in the channel, as shown in Fig.~\ref{fig:states}.
In the low state, we can either stay in this state by sending ``0'' or go to high state by sending ``1''. In the high state, however, for sending the bit ``1'' again, it is not efficient to wait for the channel to reset to ``L'' state. Instead, the transmitter can send ``1'' which is shorter in time duration and needs less molecules or equivalently less energy. To send ``0'' when channel is at state ``H'', it suffices that the transmitter waits a specific amount of time such that the channel clears itself from the molecules; returning to ``L'' state.

The above discussion implies that depending on the channel state, duration of bit ``0'' is different. Likewise, the duration for which the transmitter emits molecules for bit ``1'' depends on the channel state.
Therefore, effectively, we can think of four different symbols sent by the transmitter, namely, $s_0=$``00'', $s_1=$``01`'', $s_2=$``10`'', and $s_3=$``11`''.
The first bit in each symbol indicates the previous bit which was already sent (and hence the channel state), and the second bit is the one to be transmitted next. Assuming a binary symmetric source, all these four symbols are equiprobable. However, a different transmission time is associated with each, as we explained.

Since our model involves symbols with various time duration, we adapt the method by Shannon in~\cite{Shannon_math} to compute the capacity of a discrete noiseless channel as
\begin{equation}
\label{eq:shannon}
C=\lim_{T\to\infty} \frac{\log N(T)}{T},
\end{equation}
where $N(T)$ is the number of allowed blocks of duration $T$. Let $T_{ij}^{(s)}$ be the duration of $s$-{th} symbol which is allowable in state $i$ and leads to state $j$. Then $N_j(T)$, the number of blocks of length $T$ ending in state $j$ is given by
\begin{equation}
\label{eq:sh1}
N_j(T)=\sum_{i,s} N_i(T-T_{ij}^{(s)}).
\end{equation}

The asymptotic solution for these difference equations would be in the form $N_j=A_j W^T$ where $A_j$ is a constant~\cite{Shannon_math}. By substituting this form in~(\ref{eq:sh1}), the possible number of blocks would be equal to $\sum_j A_j W^T$ where $W$ can be found by solving the equation~\eqref{eq:eq1} below. We define a matrix $M=[\sum_s W^{-T_{ij}^{(s)}}-\delta_{ij}]$; $W$ is the largest real root of the determinant equation $|M|=0$, i.e.,
\begin{equation}
\label{eq:eq1}
|M|=\left|{\sum_s W^{-T_{ij}^{(s)}}-\delta_{ij}}\right|=0.
\end{equation}
Here, $\delta_{ij}=1$ if $i=j$ and is zero otherwise. Hence, following~(\ref{eq:shannon}), the channel capacity $C$ is obtained as
\begin{eqnarray*}
C &=& \lim_{T\to\infty} \frac{\log \left(\sum_j A_j W^T\right)}{T}\\
	&=& \lim_{T\to\infty} \left(\frac{log W^T}{T}+\frac{\log\sum_j A_j }{T}\right)\\
	&=& \log W.
\end{eqnarray*}
Rewriting~(\ref{eq:eq1}) for the diffusion channel described above results in
\begin{equation}
\label{eq:eq2}
\left | 
\begin{array}{cc}
W^{-T_{00}}-1 & W^{-T_{01}}\\
W^{-T_{10}} & W^{-T_{11}}-1
\end{array} \right|=0.
\end{equation}
Here, $T_s$ where $s \in\{00,01,10,11\}$ is the duration of the symbols. This leads to 
\begin{equation}
\label{eq:eq3}
W^{-(T_{01}+T_{10})}+W^{-T_{00}}+W^{-T_{11}}-W^{-(T_{00}+T_{11})}=1.
\end{equation}
By solving~(\ref{eq:eq3}) with respect to $T_{s}$ for $s\in{00,01,10,11}$, we obtain the capacity of the channel in bits per second. In the next section, we use the diffusion channel property to solve for $T_s$.

\section{Main Results}
\label{sec:main}

Consider the scenario in which for symbol $s$, the transmitter sends a pulse-shaped rate of molecules with amplitude $F_s$ to the receiver for duration $T_s$. 
It is clear that $F_s$ is zero for symbols $s_1=00$ and $s_2=10$ because the transmitter does not need to emit any molecules. Instead, it must wait specific amounts of time. We assume $F_{01}$ (for $s_1$) to be equal to $F$, the maximum rate that the transmitter can produce. For $s_3=11$, there is no need for the transmitter to send as many molecules as for $s_1$. Hence, we assume $F_{11}= \alpha F$. Thus a fraction $\alpha$ of the maximum rate is allocated to $s_3$. We denote by $c_s(r,t)$ the response at the receiver, i.e. $x=r$, at time $t$ due to transmission of symbol $s$. Using~(\ref{eq:eq7}), we have
\begin{equation}
\label{eq:eq77}
c_s^*(r,t) =\int^{T_s}_0 F_s(\tau)\frac {1}{4\pi D(t-\tau)} \exp{\left( -\frac{x^2}{4D(t-\tau)}\right)} \,\mathrm{d} {\tau}.
\end{equation}

The diffusion equations of the channel do not set any limits on $T_{00}$. It depends on the receiver sensitivity, the distance $r$ between the transmitter and receiver and the diffusion coefficient $D$.  Specifically, $T_{00}$ can be considered as the time sensitivity that the receiver can sense the medium. Hence, $T_{00}$ is the fundamental time. The other $T_s$ for $s \in \{01,10,11\}$ can be considered as a multiple of $T_{00}$. Based on the Markov chain model, we derive the equations for $T_s$ for $s \in \{01,10,11\}$ corresponding to different symbols.

Assume channel is in the low state. The transmitter may send bit ``0'', i.e $s_0=00$, or change the channel state to high, i.e. $s_1=01$. In the former case, the transmitter must wait for an interval of $T_{00}$ but in the latter, it must emit molecules into the channel. Hence, for the $s_1=01$, we will have (for $ 0\leq t\leq T_{01} $)
\begin{equation}
\label{eq:eq8}
c^*(r,t) = c_{01}(r,t)
.\end{equation}

\input{analysis}

\bibliographystyle{IEEEtran}
\bibliography{ISIT11_capacity}

\end{document}

%% file: intro.tex
Our interest to study capacity of molecular diffusion channels is twofold. First, there has been numerous evidence of the existence of different forms of communication in nature. Communication enables single cells to process sensory information about their environment (in a way similar to neural networks) and evaluate and react to chemical stimuli.  At the microorganism scale, molecular signals are used for communication and control among cells. For example, one of the well-known communication primitives among cells is the phenomenon called \emph{Quorum Sensing}. Quorum sensing is a cell-to-cell communication process in which bacteria use the production and detection of molecules to monitor the population density of the bacteria. Quorum sensing allows bacteria to synchronize the behavior of the group, and thus act as a unit~\cite{Bassler1999,Bassler2002,Hammer2003,Mehta2009}.
The communication can also appear in other forms such as \emph{calcium signaling}. 
%
%
Cells absorb calcium molecules in response to various stimuli that open/close particular channels on the cell membrane. The molecular information in the variation of the calcium ions concentration is propagated both inside and outside the cell, causing a variation in the electrical charge of the cell membrane and, subsequently, the transduction of the information into an electrical signal.
Due to various limitations such as size (size of a single cell or microorganism) and energy in small scales, the dominant form of communication in such scales is via molecular signals, which is fundamentally different from conventional electromagnetic-based communication.

Secondly, recent advances in bio-nano technology has motivated research on designing nanoscale devices to perform tasks similar to their biological counterparts~\cite{Akyildiz2010}. There is a large number of applications that nano-devices could apply to.  One may envision molecular based networks built using these nanoscale devices that can be deployed over or inside the human body to monitor glucose, sodium, and cholesterol levels, to detect the presence of different infectious agents, or to identify specific types of cancer. Such networks will also enable new smart drug administrative systems to release specific drugs inside the body with great accuracy and in a timely manner. To enable all such applications, the communication among nano-devices is the key.

The above promising outlook has inspired development of new theoretical frameworks for molecular communication such as~\cite{eckford, eckford2010}. These studies by Eckford, et. al. mostly focused on molecular communication, with information conveyed by the time of the release of molecules while the propagation between the transmitter and the receiver is governed by Brownian motion (without drift~\cite{Eckford2008} and with drift~\cite{eckford2010,Kadloor2009}).  An alternative molecular communication system is the one governed by diffusion process in the medium, called \emph{diffusion channel}. Arguably, the most dominating form of the communication at the micrometer scale is diffusion based molecular communication, i.e., embedding the information in the alteration of the concentration of the molecules and rely on diffusion to transfer the information to the destination. The communication among bacteria, calcium signaling, pathogen localization functionality of the immune system and many others can be reduced to diffusion based molecular communication. Another body of work in this field involves the study of the Quorum Sensing as a network and mapping the Quorum Sensing to consensus problem under diffusion-based molecular communication~\cite{CISS2011_Arash}.

\begin{figure}
\centering
\vspace{-0.05in}
\includegraphics[width = .6\columnwidth, angle=-90]{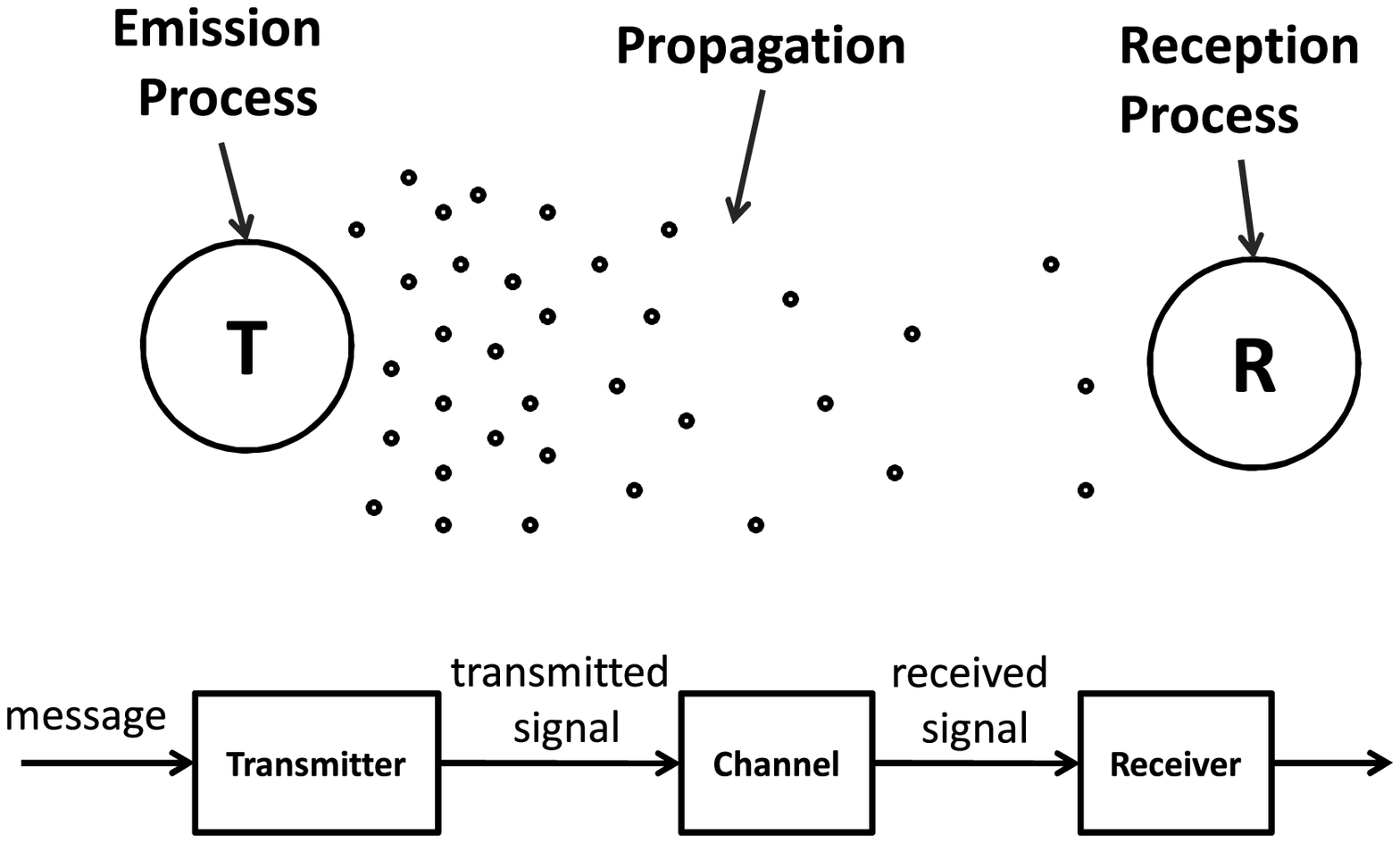}
\caption{ Scheme of the model for molecular communication}
\label{fig:model}
\vspace{-0.15in}
\end{figure}

In this work, we study the diffusion channel.
We consider a communication scenario consisted of a transmitter and a receiver communicating via the diffusion channel, as shown in Fig.~\ref{fig:model}. In this context, we can think of the transmitter as a nano-device or a bio-engineered cell capable of emitting molecules and releasing them into the medium to change their concentration according to information bits. Similarly, the receiver is capable of absorbing the molecules or chemical signals, for example, by using ligand-receptor binding which is a transmembrane receptor protein on a receiving cell~\cite{Keramidas2004,Model1995}. This ligand-receptor interaction creates peculiarities such as non-Gaussian noise. The receiver has a large number of binding places using which it can estimate the concentration by averaging over all binding places. The diffusion process has a profound impact on transmission of information that makes the diffusion channel very different from the classical models developed for electromagnetic-based communication. The molecules that are produced by the transmitter stay in the medium and affect the later transmissions. Our goal is to compute the maximum achievable rate of information exchange (hereafter referred to as \emph{capacity}) in such a diffusion based molecular communication channel. Analogous to the early studies of communication systems, we avoid some of the peculiarities of the system in our first step and we focus on studying the simpler and more practical \emph{discrete noiseless} systems. According to Shannon, in a discrete communication system, a sequence of choices from a finite set of elementary symbols $\mathcal{S}$ can be transmitted from transmitter {\bf{T}} to receiver {\bf{R}}. Each symbol $s_i$  is assumed to have duration $T_{s_i}$. As it it will be discussed later, it is important to notice that the symbols have different durations imposed by the diffusion process.

Therefore, to compute the capacity, we first characterize, the duration of each symbol by a careful study of diffusion.

In Sec.~\ref{sec:back} we review some results about diffusion process. In Sec.~\ref{sec:model} we introduce the communication model and investigate the problem under study in more detail. Our main results are discussed in Sec.~\ref{sec:main}.

%% file: analysis.tex
On the other hand, when the channel is in the high state, we have to take into the account the effect of the existing molecules in the channel, i.e. the memory of the channel. We can replace the molecules in the channel by the source that produces this concentration before the transmission of the next symbol and is zero afterwards. Based on our model, we can capture the memory of the channel by incorporating the effect of the symbol $s_1$ on the next symbol. In the case that the transmitter wants to send ``0'' afterwards, i.e. $s_2=10$, it must wait for the channel to become low. Hence, we have (for $ 0\leq t\leq T_{10} $)
\begin{equation}
\label{eq:eq9}
c^*(r,t) =c_{01}(r,t+T_{01})+c_{10}(r,t)=c_{01}(r,t+T_{01}).
\end{equation}
where the first term corresponds to the effect of the memory. The second term $c_{10}(r,t)$ is equal to zero which implies that the transmitter does not need to send any molecules but has to wait an appropriate amount of time. Finally in the case that the transmitter wants to send ``1'' again, i.e. $s_3=11$, we have (for $ 0\leq t\leq T_{11} $) 
\begin{equation}
\label{eq:eq10}
c^*(r,t) = c_{01}(r,t+T_{01})+c_{11}(r,t).
\end{equation}
Here the first term in~(\ref{eq:eq10}) corresponds to the effect of the channel memory and the second term is due to the emission of molecules for this symbol.

In the absence of noise, any intended concentration can reach the receiver without any interference from the medium. Therefore, the only limiting factor for capacity is the memory of the channel imposed by diffusion transmission. Let $S$ denote the concentration sensitivity of the receiver. This implies that the receiver cannot differentiate between the levels of concentration that differ by less than $S$. We map the low concentration on $S$ and high concentration on $2S$. We want the concentration at the receiver to be equal to one of these values at the end of each interval, i.e. $t=T_s$. 
Using~(\ref{eq:eq8}),~(\ref{eq:eq9}), and~(\ref{eq:eq10}), we have
\begin{equation}
\label{eq:eq11}
\begin{array}{rl}
& c_{01}(r,T_{01})=2S,\\
& c_{01}(r,T_{01}+T_{10})=S,\\
& c_{01}(r,T_{01}+T_{11})+c_{11}(r,T_{11})=2S\\
\end{array}
.\end{equation}
Note that in these equations the time delay between the transmitter and the receiver is not considered. We can write the solution for~(\ref{eq:eq11}) based on the function $\text{Ei}(x)$ defined as~\cite{Milton1964}
\begin{equation}
\label{eq:eq12}
\text{Ei(x)}=\int^\infty_x {\frac{\exp(-y)}{y}} \,\mathrm{d}y
.\end{equation}
Then expressions in~(\ref{eq:eq11}) can be written as
\begin{equation}
\label{eq:eq13}
\begin{array}{rl}
& \text{Ei}\left(\frac{R^2}{4 D T_{01}}\right)=\frac{8\pi DS}{F},\\
& \text{Ei}\left(\frac{R^2}{4 D (T_{01}+T_{10})}\right)- \text{Ei}\left(\frac{R^2}{4 D T_{10}}\right)=\frac{4\pi DS}{F},\\
& \text{Ei}\left(\frac{R^2}{4 D (T_{01}+T_{11})}\right)- (1-\alpha) \text{Ei}\left(\frac{R^2}{4 D T_{11}}\right)=\frac{8\pi DS}{F}.\\
\end{array}
\end{equation}

Since finding closed form formula for (\ref{eq:eq13}) is not possible, in the following we solve for (\ref{eq:eq13}) numerically to obtain $T_{01}$, $T_{10}$, and $T_{11}$ in terms of $F$. The parameter $F$ indicates the maximum power used by the transmitter which controls the capacity. 
We solve~(\ref{eq:eq13}) based on the normalized parameter $\tilde{F}=\frac{F}{4\pi DS}$ and obtain $T_s$ for $s \in \{01,10,11\}$ as multiples of $\frac{R^2}{4D}$. As discussed before, $T_{00}$ depends on $r$ and $D$. Therefore, we may assume $T_{00}=\frac{kR^2}{D}$ where $k$ is a constant. We arbitrarily choose $k=1$, although our analysis holds for any other values of $k$. 

\begin{figure*} 
\centering 
\vspace{-0.15in}
\subfigure[]{
\includegraphics[width=2.4in]{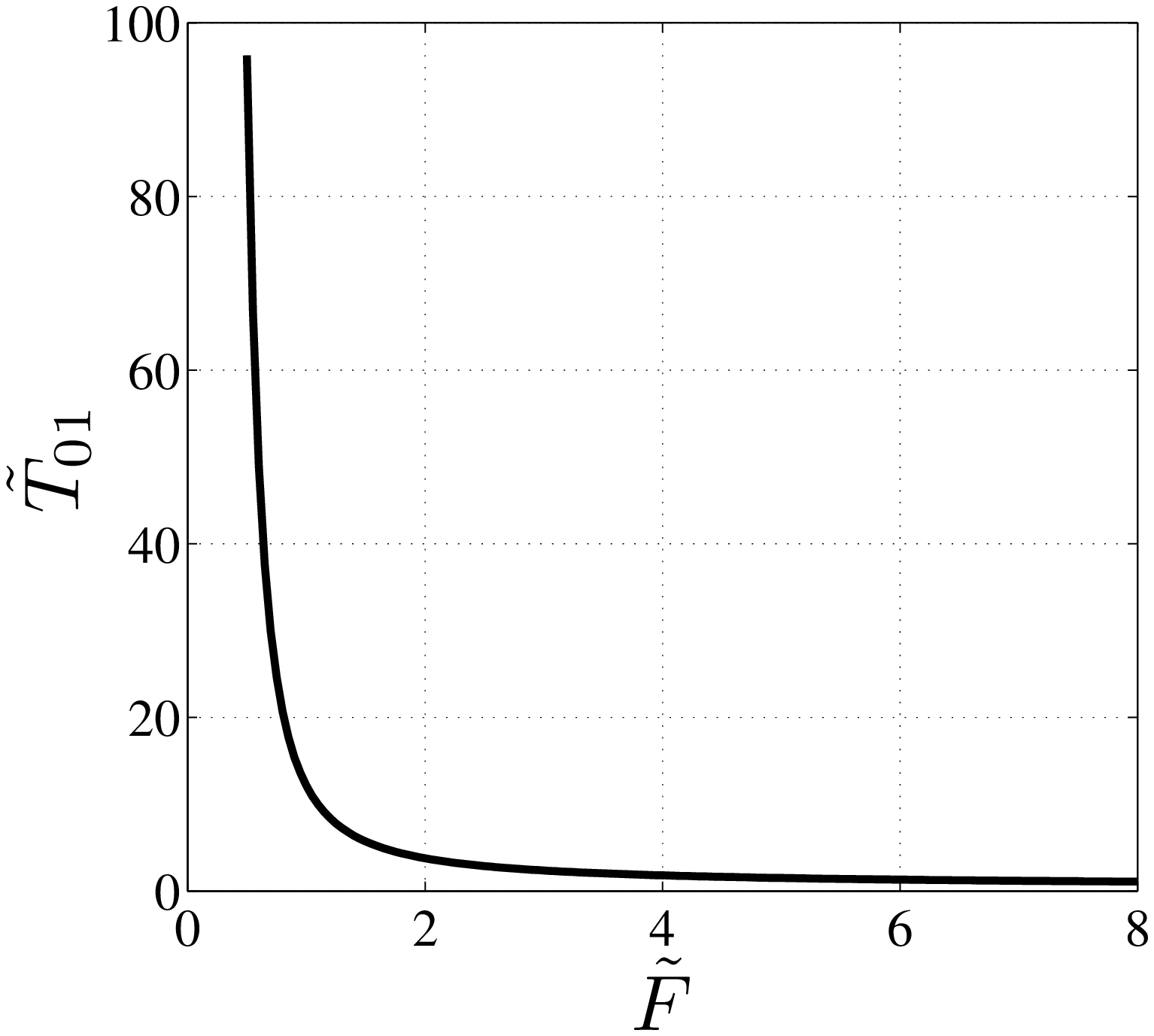}
\label{fig:fig4}
} 
\subfigure[]{ 
\includegraphics[width=2.4in]{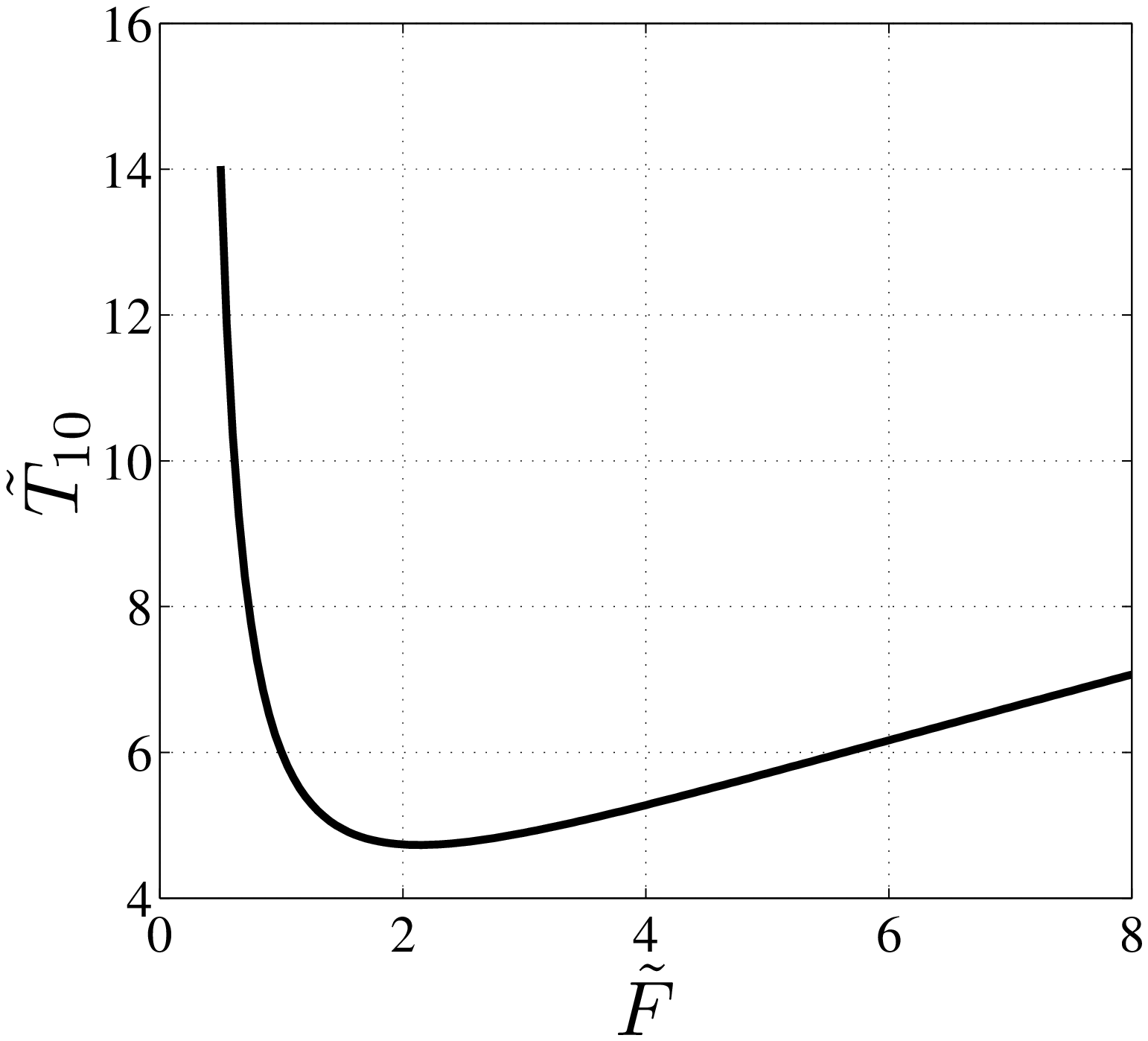}
\label{fig:fig5}
} \caption{Duration of time needed for transition from a) low to high, and b) high to low versus maximum rate of production.}
\vspace{-0.15in}
\label{fig:transition}
\end{figure*}

It can be shown that $T_{11}$ is smaller, by orders of magnitude, than both $T_{01}$ and $T_{10}$. Intuitively, this is due to the fact that producing a high state needs much more molecules than maintaining it. Further, as indicated in the following, $T_{01}$ and $T_{10}$ are in the same order of magnitude for typical values of $F$. Hence, for simplicity, we assume $T_{11}$ to be equal to $T_{00}$, and perform our numerical analysis only on $T_{01}$ and $T_{10}$. The numerical results for $\tilde{T}_{01}=\frac{T_{01}}{T_{00}}$ and $\tilde{T}_{10}=\frac{T_{10}}{T_{00}}$ are shown in Fig.~\ref{fig:fig4} and~\ref{fig:fig5} respectively.
%
%

\begin{figure}
\centering
\vspace{-0.05in}
\includegraphics[width = 2.4in]{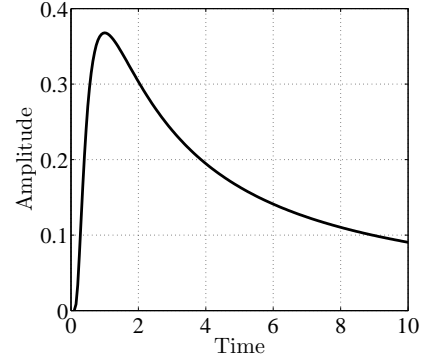}
\caption{Impulse Response of the Diffusion Channel.}
\vspace{-0.15in}
\label{fig:impulse}
\end{figure}

Fig.~\ref{fig:transition} shows the plots of transition time versus the normalized maximum rate of the molecule production $\tilde{F}$. We observe that the transition time from low to high $\tilde{T}_{01}$ is monotonically decreasing with respect to $\tilde{F}$, and approaches to infinity when $\tilde{F}$ is close to zero.
%
In Fig.~\ref{fig:fig5}, the normalized transition time from high to low state $\tilde{T}_{10}$ is presented. It has a minimum around $\tilde{F}=2$. This behavior can be explained by the diffusion equation as following. Using~(\ref{eq:eq5}), the impulse response of the diffusion channel is shown in Fig.~\ref{fig:impulse}.
As we see in this plot, the impulse response increases to a maximum and then falls lower. To obtain the response to the transmitter, a pulsed shape signal in time has to be convolved with the impulse response. When we increase the $\tilde{F}$ (or equivalently decrease $\tilde{T}_{01}$), at some point the threshold $2S$ is reached before the maximum of impulse response. Since the tail of the impulse response decreases slowly, in order to send the symbol ``10'', the transmitter has to wait until the pulse is slided to the right side of the maximum. When $\tilde{F}$ is smaller and $\tilde{T}_{01}$ is larger, we will not face this problem. 
Therefore, using Fig.~\ref{fig:fig5}, we conclude that increasing the rate of production $\tilde{F}$ does not always result in smaller $\tilde{T}_{10}$. Although larger $\tilde{F}$ results in smaller $\tilde{T}_{01}$, it makes $\tilde{T}_{10}$ to grow larger, and hence has a negative effect on the capacity. Hence, the optimum value for the production rate can be found by trade-off between $\tilde{T}_{01}$ and $\tilde{T}_{10}$.

\begin{figure}
\centering
\vspace{-0.05in}
\includegraphics[width = 2.5in]{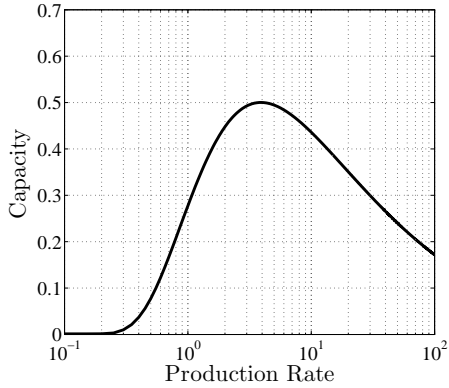}
\caption{Capacity vs Production Rate.}
\vspace{-0.15in}
\label{fig:capacity}
\end{figure}

Using~(\ref{eq:eq3}) and the assumption that $T_{11}=T_{00}$, the capacity of the channel can then be obtained by solving for $W$ in the following equations.
\begin{equation}
\label{eq:eq15}
W^{\tilde{T}_{01}+\tilde{T}_{10}}-2W^{\tilde{T}_{01}+\tilde{T}_{10} -1}+W^{\tilde{T}_{01}+\tilde{T}_{10} -2}=1,
\end{equation}
and using it in $C=\log W$. The capacity, $\log W$, is in  ``bits per an interval of $T_{00}$''. We note that maximum capacity is achieved when $\tilde{T}_{01}+\tilde{T}_{10}$ is minimized. The plot of the resulting capacity is shown in Fig.~\ref{fig:capacity} with respect to $\tilde{F}$. 
As shown in the figure, the maximum capacity is approximately 0.5 (bit/$T_{00}$) which is achieved when choosing $\tilde{F}=3.9$ (units) and corresponds to the minimum value of $\tilde{T}_{01}+\tilde{T}_{10}$ which is approximately 7 (time units). The left tail of the graph converges to zero which corresponds to the no production rate. The right tail of the graph converges to zero as well but at a slower pace. In this case, symbols ``00'', ``11'', and ``01'' can be transmitted assuming the channel is in the proper state. However, since $\tilde{T}_{10}$ becomes very large, the transition between the channel states takes a long time, leading to zero capacity.
\vspace{-.15in}
\section{Conclusion}
In this paper, we used the discrete noiseless channel model to obtain the capacity of the diffusion channel. The inherent memory in the diffusion channel distinguishes it from typical wireless channels, making the capacity analysis more challenging. For a binary input, we considered four different symbols to take into account the effect of the channel state on the next transmitted bit. With a numerical analysis, we showed that beyond a threshold, increasing the level of power has a negative effect on the capacity, which was explained by the physical characteristics of the diffusion channel.